# Strain analysis of Ge micro disk using precession electron diffraction

Aneeqa Bashir,[1] Ross. W. Millar,[2] Kevin Gallacher,[2] Douglas. J. Paul,[2] Amith. D. Darbal[3], Robert Stroud,[3] Andrea Ballabio,[4] Jacopo Frigerio,[4] Giovanni Isella,[4] and Ian MacLaren[1]

[1]School of Physics and Astronomy, University of Glasgow, Kelvin Building, University Avenue, Glasgow G12 8QQ, United Kingdom

[2]School of Engineering, University of Glasgow, Rankine Building, Oakfield Avenue, Glasgow G12 8LT, United Kingdom

[3]NanoMEGAS USA, Tempe, Arizona, USA

[4]L-NESS, Dipartimento di Fisica del Politecnico di Milano, Polo Territoriale di Como, Via Anzani 42, Como I-22100, Italy

**ABSTRACT**

The recently developed precession electron diffraction (PED) technique in scanning transmission electron microscopy (STEM) has been used to elucidate the local strain distribution and crystalline misorientation in CMOS fabricated strained Ge micro disk structure grown on Si substrate. Such structures are considered to be a compact optical source for the future photonics due to the specific undercut for direct bandgap behaviour under strain. In this study, the strain maps are interpreted and compared with a finite element model (FEM) of the strain in the investigated structure. Results demonstrate that the SiN used as a stressor on top of the Ge disk induces an in-plane strain $\varepsilon_{xx}$ of a maximum value of almost 2 % which is also confirmed by FEM simulations. This tensile strain can reduce the difference between the direct and indirect bandgaps leading to direct bandgap radiative transitions, with the potential for applications in strained Ge lasers.

1. **INTRODUCTION**

It is only recently that Ge has made ground in standard Si technology for being compatible with CMOS processing [1]. Ge on Si has been established as a low cost photonic component with promising applications including photodetectors [2] and modulators. [3, 4] In the last decade, Ge has also been investigated for active optical sources, and lasers. This has been driven by the fact that, despite being an indirect bandgap material, the direct $\Gamma$-valley is only 140 meV above the L-valley, meaning that electrical population of the direct valley is possible. Lasing has been demonstrated in Ge Fabry-Perot cavities with high degenerate n-type doping, and low tensile strain from growth on Si [5-7]. This doping reduces the required injection to populate the $\Gamma$, thereby minimising the excess injected hole population, and the associated strong inter-valence and free carrier absorption (FCA). Degenerate phosphorous doping, however, leads to a reduced excess carrier recombination lifetime [8] and still leads to free carrier losses, which produces high lasing thresholds and has limited the number of demonstrations of room temperature Ge/GeSn lasing with low strain.

Applying biaxial tensile strain to Ge causes the $\Gamma$-valley to lower in energy at a greater rate than the L-valley, leading to a cross over to a direct band-structure at ~ 1.7 % [9]. This reduces the required injection to populate the $\Gamma$-valley, and therefore reduces FCA/inter-valence band losses. Two major approaches have been used for strain engineering Ge. Locally undercut photonic wires or microbridges can locally amplify the low tensile strain from the epitaxial growth of Ge on Si [10], resulting in large uniaxial strains of up to 5.7 % [11] at room temperature. Lasing has been reported at low temperatures by this approach [12]. SiN stressors have also been used to externally apply strain to compact micro-disk structures, with sufficient biaxial tensile strain to transition to a direct bandgap bandstructure[9, 13]. Recently, lasing at low temperatures has been demonstrated in highly tensile strained Ge micro-disks[14].

Alloying Ge with Sn has also emerged as a viable means for creating a direct bandgap group IV material. Similarly to tensile strain, increasing Sn content causes a transformation to direct-bandgap[15]. There have now been multiple demonstrations of lasing with GeSn[16-18] up to a record high temperature of 230 K, with Sn concentrations ~ 16 % [19]. Despite the huge success of GeSn alloying, a moderately direct band-structure is not sufficient for room temperature lasing, as electrons can thermally escape the $\Gamma$-valley and scatter into the L-valley, which has a much higher density of states. The $\Gamma$-valley has to be significantly lower in energy (several times $k_BT$) to prevent this effect. This is problematic as the high Sn concentration layers required for such a $\Delta E_{L-\Gamma}$ result in challenging material growth, low thermal stability, and a high defect density that reduces the carrier lifetime.

It is therefore likely that a combination of external strain engineering and Sn alloying is required to achieve room temperature lasing, in order to improve the level of 'directness' i.e. $\Delta E_{L-\Gamma}$, while keeping thermal budgets sufficiently high for Si foundry processing, and potentially improving gain due to the strain induced splitting of the valence band[20-22]. The use of SiN stressors for strain engineering GeSn is highly applicable[23, 24], as the microbridge approach is not feasible with layers that are compressively strained as grown.

To fully understand the potential modal gain available in SiN strain engineered Ge/GeSn devices, it is crucial to have an accurate measurement of the strain field in the devices cross section, with high spatial resolution. This is not achievable with measurements such as Raman spectroscopy, which only probes the top plane, averaged over the penetration depth of the excitation source. As an alternative, the use of the transmission electron microscope (TEM) provides an instrument in which spatial resolutions of a few nm are routinely achieved, and in which diffraction-based techniques could provide suitable tools for strain measurement and spatially-resolved mapping. Here, we demonstrate the use of a Precession Electron Diffraction (PED) technique to produce a cross-sectional strain map of a SiN strained Ge micro-disk, and compare results to FEM of the strain field. The results are in good agreement with the modeling, and the previously measured optical properties of the structure, but also reveal the strain field around threading dislocations, and the resulting deviations from the simplified model, which does not include defects and dislocations.

**High Spatial Resolution Strain Measurement in the Electron Microscope**

Strain measurements are critical to understand and optimise the properties of the materials for improved device applications. It is therefore indispensable to have a method that can quantify the strain over nanoscale resolution and with high precision. A number of methods to investigate strain in semiconductors have been employed. Early studies used X-ray diffraction [25, 26] and Raman spectroscopy.[27-29] These techniques are well established now and routinely have a spatial resolution of hundreds of nanometres.

As an alternative, the use of the transmission electron microscope (TEM) provides an instrument in which spatial resolutions of a few nm are routinely achieved, and in which diffraction-based techniques could provide suitable tools for strain measurement and spatially-resolved mapping. Béché *et al.* have recently compared some of the methods for strain mapping in TEM[30], and these will be briefly summarised below.

Geometrical phase analysis (GPA) of atomic-resolution TEM or scanning TEM (STEM) images [31-33] provides good spatial resolution but poor precision.[34, 35] The issue with using GPA is that to identify the lattice changes correctly because lattice strain phase shifts are used, and changes in relative intensity of the sublattices and sub unit cell displacements can all also produce phase shifts in Fourier components[36] of TEM images, which can easily be misinterpreted. Additionally, good atomic resolution images are only possible for very thin specimens, and in structures (such as the case studied in this research) where there is a large biaxial strain, then strain relaxation in the out-of-plane direction could mean that the measurements in the thin specimen bear no simple relation to the strain state of the original structure.

Convergent beam electron diffraction (CBED) [37] has been successfully used to make strain measurements in semiconductor devices. The most important limitation of this methodology for strain analysis in crystal materials, however, is that reliable pattern simulations are required to fit with experimental results, which are time-consuming and model-dependent.[38] This does not suit when large volumes of specimens are to be examined, especially in semiconductor research, and is not necessarily suitable for the mapping of strain fields.

Dark-field electron holography (DFEH) was first employed by Hÿtch *et al.*[39] for strain measurement in electronic devices in 2008. Geometric phase measurements of the hologram (an interference pattern obtained from strained and unstrained region) were employed to map the strain at the nanoscale [30, 31, 40], which allows maps to be calculated with a high spatial resolution of around 5 nm and a precision of ±0.02%. [35, 39] A major difficulty with this technique, however, is that the measurements at the region of interest are performed at an angle to a major zone axis, and this can be problematic in thin layers. Other restrictions are that the sample should have a uniform thickness all around [41] and an unstrained reference region must lie next to the region of interest, which is not always possible in real devices [42].

In nanobeam electron diffraction (NBED), local diffraction patterns at each point are acquired by scanning an electron probe (~5−10 nm) across a specimen. These diffraction patterns are compared with a reference pattern in order to provide relative shifts in the diffraction spots to quantify the deformation in the specimen.[43-45] Though a precision of roughly ±0.1 % can be achieved with this technique, results can be erroneous and noisy at times. This occurs due to the contrast within diffraction spots arising from multiple scattering processes (dynamical effects).[46] Also, to improve the spatial resolution, a larger convergence angle is required, which makes the diffractions spots into large discs which are even more prone to dynamical scattering within the discs. To overcome these hurdles, a method was developed of precessing the electron beam at each point on the specimen whilst acquiring the diffraction pattern.[47] Thus, the intensity at each diffraction spot is integrated from many incident angles, which averages out most of the dynamical effects within the spots, creating small discs of uniform intensity, allowing their automated detection and locations with subpixel accuracy. Most importantly, precession also enables the collection of higher-order reflections, which are more sensitive than lower order reflections to small changes in lattice parameters. This technique is known as Scanning Precession Electron Diffraction (SPED). For an application like the present one, this is the ideal technique – it can be applied in relatively thick material (~ 100 nm), has a resolution of a few nm (mainly limited by beam spreading in a thicker sample), can be performed along a major zone axis, and has high precision with a sufficiently large precession angle to include higher index spots.[48, 49] Though the use of SPED for strain measurement is relatively new, it has already been employed to study strain in SiGe and Ge based semiconducting nanostructures and devices.[48-53]

In this paper, we report on high precision SPED measurements to accurately map the strain in a micro-disk shaped Ge structure in which the strain is concentrated in a small region. These experimental measurements are then compared with Finite Element Modelling (FEM) to evaluate the strain, and the differences are highlighted, and the microstructural reasons for these differences are investigated using conventional TEM techniques. Our work demonstrates the high potential of the SPED technique for the investigation of inhomogenously strained microstructures with a prospect of applications in future photonics.

## 2. EXPERIMENTAL DETAILS
### 2.1 Fabrication of Ge- microdisk Structure

Low-Energy Plasma-Enhanced Chemical Vapor Deposition (LEPECVD) was used to grow a 380 nm of Ge heterostructure on top of a Si (001) substrate at low temperatures of 500˚C.[54] The Ge structure was patterned by electron beam lithography in a Vistec VB6 tool using hydrogen silsesquioxane (HSQ) resist and was dry etched afterwards in a mixed $SF_6$ and $C_4F_8$ recipe through to the Si layer.[55] Si was anisotropically wet etched to undercut the structure to leave a controlled size of Si post to support the disk, using tetramethylammonium hydroxide (TMAH) and isopropyl alcohol (IPA). This was followed by coating of the structures with high stress silicon nitride in an inductively coupled plasma enhanced chemical vapour deposition (ICP-PECVD) tool. This produces a high compressive stress of ~2.4 GPa in the film, controlled through the deposition parameters [56]. More details of growth can be found in Millar *et al.*[13].

### 2.2 Microscopy specimen preparation

A cross section sample was prepared using a focussed ion beam (FIB, FEI Nova Nanolab 200) lift out procedure[57, 58] to provide a thin slice (lamella) of suitable and even thickness for analysis by STEM. After identification of a suitable area for cross-sectioning a disc, a thin platinum (Pt) layer was deposited using electron beam deposition to protect the TEM specimen cross-section from the $Ga^+$ ion beam induced damage (this should not stress the structure any further). Trenches were then created on both sides and parallel to the structure to create a 1μm thick and 15μm wide lamella. This was carefully lifted out in situ using the micromanipulator tip and attached to an Omniprobe support grid. Finally, FIB milling and polishing were carried out on both sides of TEM lamella, using a reduced beam current, to produce a nearly parallel sided specimen of final thickness of about 100 nm (the sample was left a little thicker than some (S)TEM specimens used for atomic resolution analysis, in order to leave more bulk-like strain states within).

### 2.3 Precession electron diffraction and Imaging

SPED was performed using a 200 kV CM200 (S)TEM equipped with a field emission gun (FEG) operated in the STEM mode and using a convergence half angle of 0.7 mrad. Precession of the electron beam was performed at an angle of 1.8º and a frequency of 100 Hz using a NanoMEGAS DigiSTAR unit. The probe size was ~10 nm for the chosen condenser aperture of 20 μm. The reference diffraction pattern, obtained from an unstrained region, was modified numerically to make it comparable with each diffraction pattern taken from a strained region. This does not require identifying individual diffraction spots and thus make it easier to derive strain coefficients from processing these diffraction patterns. Thus, 2D maps for in-plane strain $\varepsilon_{xx}$, out-of-plane strain $\varepsilon_{yy}$, and shear strain $\varepsilon_{xy}$ are obtained. In order to proceed with crystalline orientation identification, thousands of simulated electron diffraction spot patterns (so called templates) were utilized for each crystallographic phase in the sample. Local crystal orientations were obtained by comparing experimentally acquired patterns using cross-correlation matching techniques with the template database.

STEM imaging of the structure was undertaken on a probe corrected JEOL ARM 200F equipped with a cold field emission gun operated at 200 kV in STEM mode and using a convergence half angle of 29 mrad. High angle annular dark field (HAADF) imaging was performed with an inner detector angle of 107 mrad. At this angle the contrast is strongly dependent on atomic number and therefore the image brightness is strongly correlated to chemistry. A Gatan GIF Quantum ER energy filter/spectrometer with fast Dual EELS (electron energy loss spectroscopy) was used to record the spectrum images (SIs) with a collection half angle of 36 mrad and a dispersion of 1.0 eV per channel, to determine the final thickness of the lamella using the *t/λ* method. Dark field TEM imaging was performed on a FEI Tecnai T20 operated at 200 kV in order to view dislocations in the Ge micro-disk and in an equivalent specimen constructed from a Ge heterolayer grown in an identical way but not fabricated into a strained micro-disk.

### 2.4    Finite Element Simulations

Finite element simulations[59] have been used to generate strain maps which have been compared with experimental results. A three-dimensional FEM model was employed in COMSOL Multiphysics, taking into account the anisotropy of the Ge elasticity tensor and using the geometry of the 4 μm micro-disk measured by STEM (shown in Figure 1). The model also includes the 2.45 GPa silicon nitride stressor, which was optimised to make the vertical displacement of the micro-disk edge in the model comparable to that measured by STEM in Figure 1. The vertical deflection at the micro-disk edge is ∼150 nm. The model has the **x** and **y** directions orientated along the <100> crystallographic directions, which also represent the edge of the micro-disk.

### 3.    RESULTS AND DISCUSSIONS

**Strain mapping**

Figure 1 shows an overview image of a FIB section through a 4 μm diameter Ge micro-disk acquired using HAADF STEM. Different layers in the sample can be distinguished by the layer brightness and are labelled accordingly in the figure. The image shows a high-quality cross section with no "curtaining" or preferential milling artefacts, and the outer wings are clearly slightly bent, as expected. Experimental measurements using SPED, have been performed in the outer parts (wings) of the crystalline Ge; marked red in Figure 1. The axes chosen are drawn and the **z**-axis is the direction of electron beam and lies normal to the image.

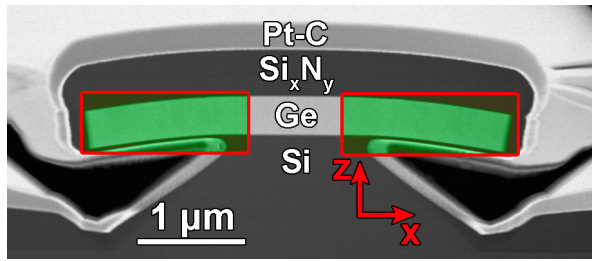

**Figure 1:** A HAADF STEM image of a 4 µm Ge micro-disk structure grown on a Si post, with SiN layer as a stressor on top.[13] Protective Platinum layers deposited during cross sectioning in the FIB are also visible. Rectangular areas marked red have been used for strain mapping using PED. The coordinate system chosen is also shown.

Figure 2 shows the experimental strain maps acquired by PED for both wings for in-plane $\varepsilon_{xx}$, out-of-plane $\varepsilon_{zz}$, and shear $\varepsilon_{xz}$ components. The lattice strains were originally evaluated with respect to the Si lattice parameter measured on the pedestal and therefore includes the lattice misfit of ~ 4.2% that exists between Ge and Si, but are corrected to refer to strains according to the lattice parameter of unstrained Ge. The red areas represent positive strain relative to the blue areas (Si). At the top of the Ge the $\varepsilon_{xx}$ strains are tensile of just below 2% but the $\varepsilon_{zz}$ values in the same regions are slightly compressive, about 1% below the Ge lattice parameter. At the base of the wings next to the post, this trend is reversed and the $\varepsilon_{xx}$ is compressive and the $\varepsilon_{zz}$ strain is tensile. Away from the post, there are smaller variations in $\varepsilon_{zz}$ but the trend of $\varepsilon_{xx}$ being tensile at the top and compressive at the bottom continues. There are also some sharp changes in the vertical strain seen in straight diagonal lines in Figures 2c and 2d. Both of these radiate from a point close to the corner of the post and suggest a breakdown in the idealised strain state that would be predicted by continuum elastic considerations. Moreover, the decrease in the strain observed along a vertical line from top to bottom of the micro-disk depends upon the diameter of the post and can vary accordingly.[60]

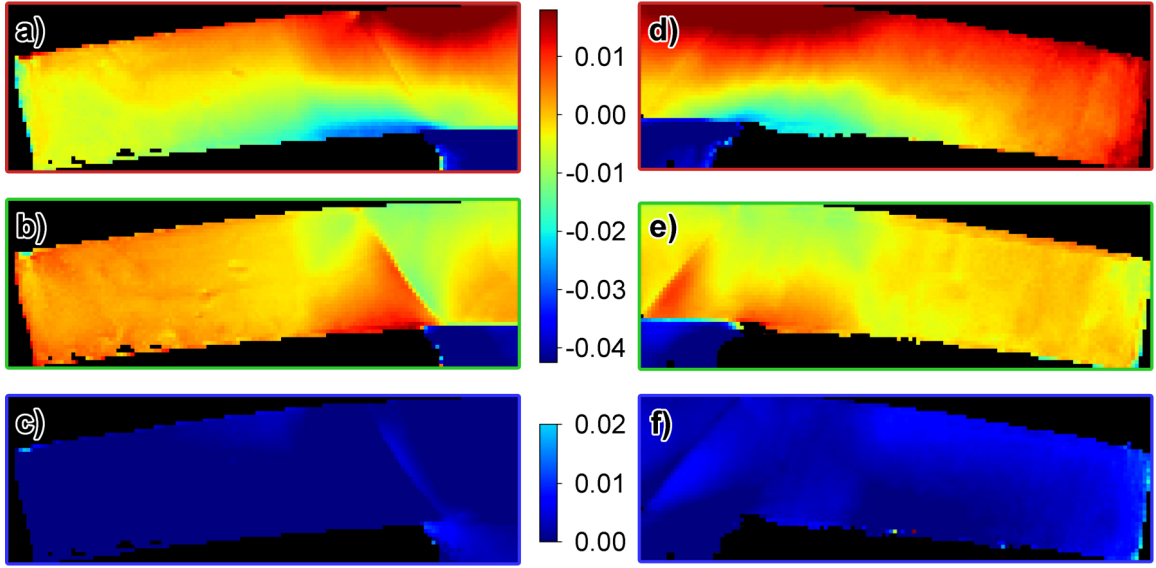

**Figure 2:** (a−c) Strain maps acquired by precession diffraction for the $\varepsilon_{xx}, \varepsilon_{zz}, \varepsilon_{xz}$, in the left wing of the Ge micro-disk, and (d-f) represents the same maps extracted for the right wing of the micro-disk. The circles and dashed lines are a guide for the eyes.

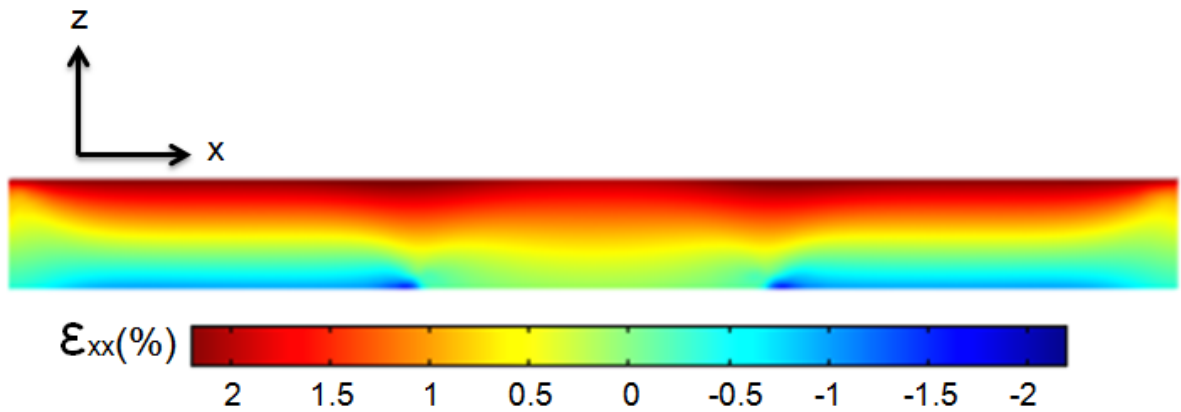

**Figure 3:** 2D slices from finite element models through a Ge micro-disk demonstrating the in-plane strain component $\varepsilon_{xx}$; it can be observed that the uniaxial stress transfer induces more strain at the micro-disk top. The geometry chosen in the model is also shown.

The best fit FEM of the in-plane strain $\varepsilon_{xx}$ in the Ge micro-disk structure is shown in Figure 3, simulated with reference to Ge and so it does not include the lattice mismatch between Ge and Si, although the meaning of the colours in absolute terms is about the same for the two maps. It shows a maximum value of ~2.1% at the top. This is slightly above the experimental measurements of the peak strain of 1.8% but certainly consistent (especially when there may have been strain relaxation in the real micro-discs, as shown below). Additionally, the simulations predict a minimum value of $\varepsilon_{xx}$ ~ - 0.5% at the bottom near the interface and this is in close agreement with experimental value of -0.6%. Raman measurements were done previously on this sample to investigate the strain components.[13] Raman measurements were consistent with an in-plane strain $\varepsilon_{xx}$ of > 2% which also agrees with the experimental and simulation results presented here. Thus, it is confirmed that the SiN stressor produces a strain of the appropriate magnitude to explain the optical and mechanical properties demonstrated in our previous work.[13] The experiments, however, only show the peak stress just outside the radius of the post, whereas simulations show it extending almost to the edge. This probably happens due to stress relaxation within the micro-disc due to plastic deformation. This means that the sample would have a ring-shaped region of peak strain at which the effects would be most significant.

FEM has certain limitations which restrict the exact agreement of the model with the experimental results. Firstly, PED measurements were performed on a thin lamella prepared from the bulk structure. In such a case, the thinning of the sample may release internal stresses in ways that alter the results from the stress distribution in the bulk.[61] Second, the epitaxy in the real sample is far more complex than the simplified model chosen in this FEM model. Specifically, it does not include factors such as any deviation from perfect epitaxy, and does not include dislocations and plastic deformation, so this could be the cause for any discrepancy between simulated and experimental strain maps. To further investigate these effects, dark field imaging and orientation mapping were performed.

**Dark field imaging**

Dark-field images have been recorded for the specimen using two-beam conditions for various diffraction vectors to reveal dislocations. Some representative images are shown in Figure 4 where the well-spaced threading dislocations are identifiable in various parts of the Ge structure as bright lines – similar threading dislocations were also observed in nominally identical Ge heterolayers without any fabrication steps. Additionally, dense arrays of dislocations in slip planes are visible marked by arrows emanating from the Si-Ge interface close to the outside edge of the post. These clearly show that plastic deformation by dislocation slip has occurred after stress concentration around this sharp corner. This was exactly the same sample as for the strain mapping in Figure 2 (although placed the other way up into the microscope, inverting left and right), and the slip planes correspond to the sharp changes in $\varepsilon_{zz}$ in those strain maps. In some samples, cracks have been observed at the outside edge of these slip planes.

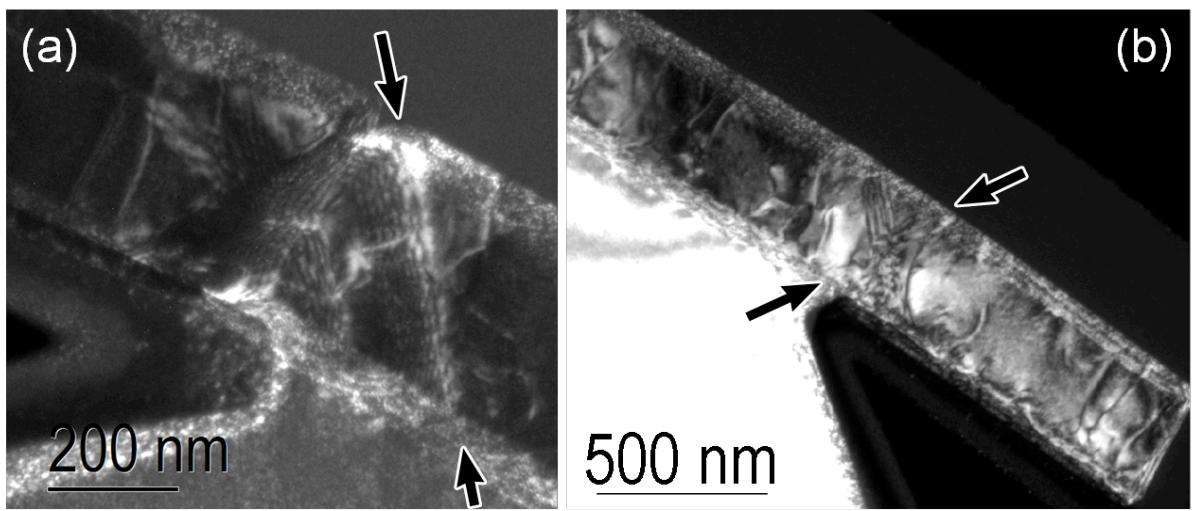

**Figure 4:** Dark field images taken across the Ge micro-disk. Bright lines are the dislocations present which can cause strain relaxation. The arrows indicate slip planes along which a larger number of dislocations moved resulting in a larger misorientation. a) detail of the right wing of Figure 2e); b) the left wing of Figure 2b). Note, the sample was placed into the microscope the other way up to in Figure 2, thus the swapping of left and right.

The threading dislocation density (TDD) was calculated by the line-intercept method[62] which has been able to provide reliable results for dislocation density measurements[63] using TEM images. In this method, randomly oriented lines are drawn through dislocations, over TEM images and the dislocations are then marked with points. The dislocation density ρ is given by the number of points N divided by the total line length of the random lines $L_r$, multiplied by the thickness t i.e., $\rho = \frac{2N}{tL_r}$. The method strictly relies upon the accurate measurement of the thickness which in the present study was found by means of electron energy loss spectroscopy using a mean free path of 97 nm calculated by the Iakoubovskii method[64] after correction by a factor of 0.8, as in previous work.[65] The mean density across the entire micro-disk was found to be $1.48 \pm 0.4 \times 10^{10}$ cm$^{-2}$. This is rather greater than that in the raw unstrained Ge grown on Si, prior to the fabrication of the disks, which is $1.6 \pm 0.5 \times 10^{9}$ cm$^{-2}$. The threading dislocation networks and density in the outer part of the micro-disk, however, appear similar to those in the unstrained material, and the main reason for the increased dislocation density in the micro-disks is slip and dislocation multiplication close to the sharp corner at the end of the undercut, as noted above. It should be noted that the Ge epitaxial layer used here did not receive cyclic annealing to reduce the TDD, to avoid diffusion and clustering of the phosphorous doping.[66] Cyclic annealing of comparable layers can be used to reduce TDD down to ~$1 \times 10^{7}$ cm$^{-2}$.[67]

**Orientation mapping**

The crystal orientation should vary with position due to the bending resulting from the SiN stressor, and this can also be investigated from the SPED datasets used for the strain mapping, as previously performed by Estradé et al..[52] The **z**-axis orientation map is displayed in Figure 5a and can be interpreted using the colour code shown. Each point, with its colour, in the map designate the crystallographic orientation of the axis of the specimen with respect to the crystal lattice, according to the colour key, which describes the orientation of the [001] direction to the vertical. The tilt angle is plotted explicitly in Figure 5b. This shows that the central portion of the micro-disk is basically oriented parallel with the silicon. The orientation changes most rapidly immediately outside the centre post. Further away from the centre post, this curvature becomes gentler, which corresponds to the reduced strains measured in the outer parts of the wings in Figure 2. Nevertheless, the overall trend in orientation with position is much as expected and corresponds with the observed stress levels, as the positions of highest stress in Figure 2 are also the areas of highest curvature in Figure 5b.

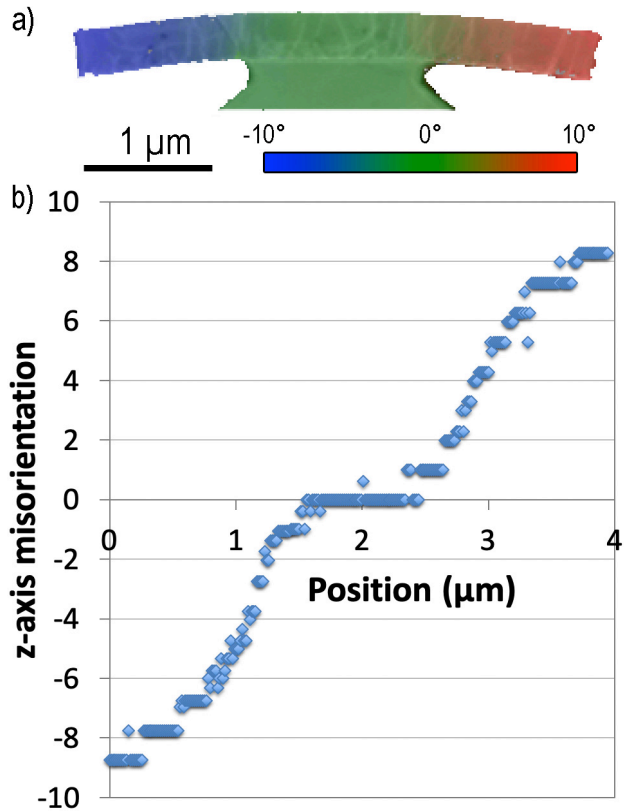

Figure 5: (a) A map of the orientation of the [001] axis to the vertical in the Ge micro-disk on Si (superimposed on the pattern quality map, which shows some variations in contrast due to dislocations), (b) the mis-orientation profile of the z-axis of the crystal structure to the vertical as a function of position going from left to right across a scan of the whole 4 μm micro-disk

**General Discussion**

In this work, we have analyzed a tensile strained Ge micro-disk structure fabricated on Si, which has been tensile strain engineered with an external SiN stressor layer. Peak strain levels are found to be consistent with experimental measurements in ref [23], which included micro-Raman and photoluminescence (PL). These structures demonstrated huge red-shifts in PL (~450 nm) and large shifts in the Raman spectra produced by the longitudinal optical phonon. Both of these techniques were consistent with a biaxial tensile strain of ~ 2 %. The discrepancies here are likely the result of some strain relaxation after the FIB process.

The strain-field results presented here are in close agreement with expectations from FEM, as was previously seen in PED-FEM comparisons of (Al,Ga)N heterostructures by Reisinger[68]. Nevertheless, there is clear evidence that some loss of elastic strain has occurred through dislocation motion and multiplication at the highest stress concentrations around the corners of the Si supporting posts. This therefore demonstrates that such techniques are extremely applicable to such strain engineered photonic structures and could be used to determine and analyze modal gain. This is particularly useful when the presence of dislocations may cause deviation from the simplified finite element models.

It is clear from this work that the single stressor layer leads to a large in-homogeneity across the micro-disk, which is undesirable for gain in Ge/GeSn structures.[69] It is, however, interesting to note that such strain engineering could be highly applicable to step graded GeSn layers. Step grading has emerged as an approach for minimizing dislocations in the active layer (highest Sn content) layer, which is grown on top of multiple layers of lower Sn content, which serve as virtual substrates. Such photonic structures have been fabricated into micro-disks and Fabry-Perot cavities and have demonstrated lasing.[16, 17, 19] In such devices, the highest Sn concentration layers have the highest refractive index, meaning that the optical mode is pulled slightly towards the top portion of the micro-disk, where the mode will overlap more with the tensile strain field. The lower, compressively strained portion of the micro-disk could comprise either Ge, SiGeSn, or low Sn content GeSn buffers that do not contribute to modal gain, and are therefore not detrimentally affected by compressive strain at the bottom of the micro-disk. In particular, higher order modes may overlap more significantly with the highest tensile strained regions of the micro-disk, however, a modal gain analysis is beyond the scope of this work.

4.  **CONCLUSION**

We have successfully characterised a Ge micro-disk structure, grown on Si by LEPECVD, using strain and misorientation mapping by scanning precession electron diffraction (SPED) with high accuracy and spatial resolution. Experimentally acquired strain maps for the $\varepsilon_{xx}$, $\varepsilon_{xy}$, and $\varepsilon_{yy}$ components have been compared with FEM simulations. Whilst the principal spatial variation of the strain and the magnitudes of peak strains match well between experiment and simulation, it is clear that some discrepancies are present. Specifically, there are sharp discontinuities in strain across diagonal bands emanating from close to the corner of the Si pillar and the Ge micro-disk above, which appear to correspond to dislocation walls along slip traces. This demonstrates that the peak strain observed here of 2.1% tensile strain at the top surface is close to the maximum that can be sustained in Ge before plastic deformation and possible cracking and failure of the structure. Nevertheless, it also demonstrates that a certain level of dislocations does not prevent the formation of sufficient tensile strains to produce the significant Raman shifts observed in our previous work.

## 5. ACKNOWLEDGMENTS

AB is thankful to the Schlumberger Foundation *Faculty for the Future* programme for the provision of a research fellowship. We are grateful to SUPA and the University of Glasgow for the provision of the JEOL ARM200F microscope used in this work. We are also grateful to Dr Stavros Nicolopoulos from NanoMegas for providing access to the scanning precession electron diffraction used in this work for the strain mapping. The work was supported by EPSRC project EP/N003225/1 and the European Union 7$^{th}$ Framework Programme project GEMINI (No. 613055).